\documentclass[aps,pre,twocolumn,superscriptaddress]{revtex4-1}
\usepackage{subfigure}
\usepackage{amsmath,graphicx,amssymb}
\usepackage{multirow,color,float}
\usepackage{url}
\usepackage{verbatim}
\usepackage{bm}
\usepackage[normalem]{ulem}

\newcommand{\vep}{\varepsilon}

\renewcommand{\vec}[1]{\bm{#1}}

\newcommand{\vu}{\vec{u}}

\newcommand{\vb}{\vec{b}}

\newcommand{\va}{\vec{a}}
\newcommand{\vk}{\vec{k}}
\newcommand{\vp}{\vec{p}}
\newcommand{\vq}{\vec{q}}

\newcommand{\vf}{\vec{f}}
\newcommand{\vw}{\boldsymbol{\omega}}
\newcommand{\avg}[1]{\langle{#1}\rangle}

\newcommand{\blue}[1]{{#1}}

\newcommand{\beq}{\begin{equation}}
\newcommand{\eeq}{\end{equation}}
\begin{document}
\vspace{-3em}
\begin{minipage}[l]{\textwidth}
Postprint version of the manuscript published in Phys. Rev. E {\bf 94}, 053209 (2016) \\
\end{minipage}
\title{{Large-scale dynamics of magnetic helicity}}

\author{Moritz Linkmann}
\email[]{linkmann@roma2.infn.it}
\affiliation{Department of Physics and INFN, University of Rome Tor Vergata, Via della Ricerca Scientifica 1, 00133 Rome, Italy}
\affiliation{SUPA, School of Physics and Astronomy, University of Edinburgh, Peter Guthrie Tait Road, EH9 3FD, UK}
\author{Vassilios Dallas}
\email[]{v.dallas@leeds.ac.uk}
\affiliation{Department of Applied Mathematics, University of Leeds, Leeds LS2 9JT, UK}

\begin{abstract} 
In this paper we investigate the dynamics of magnetic
helicity in magnetohydrodynamic (MHD) turbulent flows focusing at scales larger
than the forcing scale. Our results show a nonlocal inverse cascade of
magnetic helicity, which occurs directly from the forcing scale into the
largest scales of the magnetic field. We also observe that no magnetic
helicity and no energy is transferred to an intermediate range of scales
sufficiently smaller than the container size and larger than the forcing scale.
Thus, the statistical properties of this range of scales, which increases with
scale separation, is shown to be described to a large extent by the zero flux
solutions of the absolute statistical equilibrium theory exhibited by the
truncated ideal MHD equations.
\end{abstract}

\maketitle


The current explanation for the existence of stellar and planetary magnetic
fields is attributed to dynamo action \cite{moffatt78}.  One of the theoretical
arguments to explain the generation and preservation of magnetic fields in
spatial scales much larger than the outer scales of fluid motions is the
inverse cascade of magnetic helicity in MHD turbulence \cite{Frischetal75}.
Magnetic helicity plays a fundamental role \blue{in} 
the long-term evolution of
stellar and planetary magnetic fields \cite{brandenburgetal05} and hence its
dynamics across scales \blue{is} 
important to shed light on the saturation mechanisms
of \blue{these} 
large-scale magnetic fields. 

Previous investigations \blue{concerning} 
the inverse cascade of magnetic helicity reported
both local and nonlocal transfers with various scaling exponents \blue{measured} 
for
the spectra at large scales \cite{Brandenburg01,Alexakisetal06,Muelleretal12}.
However, it is presently unclear whether these local and non-local transfers
should be associated with a process that takes place with constant flux
\cite{Mininni11}.  Due to growing evidence for the importance of non-local
interactions in the dynamics of MHD turbulence
\cite{Brandenburg01,Debliquy05,Alexakis05a,Mininni11}, concern has been raised
over the use of the term {\em cascade} \cite{Muelleretal12,Linkmannetal16}.
Therefore, in our context the term \emph{cascade} does not necessarily imply
locality in wavenumber space.

In spite of the importance of the inverse cascade of magnetic helicity, there
is \blue{a} lack of understanding about its non-linear dynamics at large scales. In this
paper we aim to elucidate the steady-state dynamics of magnetic helicity and
\blue{its} 
role in the long time evolution of the large-scale magnetic field. To do
this we consider flows with high enough scale separation by applying helical
electromagnetic forcing at intermediate scales using \blue{direct} numerical
simulations \blue{(DNS)} and we focus on the dynamical and statistical
properties of the large scales. We show that the inverse cascade of magnetic
helicity is a manifestation of non-local transfers from the forcing scale to
the largest scales of the magnetic field in agreement with previous studies
\cite{Brandenburg01,Alexakisetal06}. Moreover, we demonstrate that despite the
fact that in three-dimensional MHD turbulence the scales between the forcing
scale and the container size are not isolated from the turbulent scales, their
statistics may still be reasonably approximated as if they were in statistical
equilibrium for high enough scale separation.

In planets and stars as well as in laboratory experiments, physical boundaries
confine fluids \blue{and set the largest possible characteristic length scale of the flow.} 
In our DNS,
the size of the periodic box $2\pi L$ is the surrogate for this spatial
confinement. In order to study the large-scale dynamics of turbulence, large
enough scale separation is necessary between the box size and the forcing scale,
while \blue{at the same time one has to ensure that}
small-scale turbulence is still resolved.  
Forcing at intermediate scales
and aiming for a turbulent flow with high enough scale separation is \blue{therefore} almost
prohibitive even with today's supercomputers.  We partly circumvent this
difficulty by considering hyper-dissipative terms under the assumption that the
dissipative scales should not significantly affect the statistical properties
of the large scales. Due to the presence of the inverse cascade of magnetic
helicity we consider a large-scale dissipation mechanism to saturate the
expected energy growth.  Otherwise, energy accumulates in the largest scales of
the box until it is balanced by viscosity leading to the formation of very
large amplitude vortices \cite{da15}.
%
Therefore, we consider the following dynamical equations: 
\begin{align}
(\partial_t -\nu^- \Delta^{-m} -\nu^+ \Delta^{n})\vec{u}&
= \vu \times \vw + \vec{j} \times \vec{b} - \nabla P + \vec{f}_u \ , 
\nonumber\\ 
(\partial_t -\eta^- \Delta^{-m} -\eta^+ \Delta^{n})\vec{b}&
=  \nabla \times (\vu \times \vb) + \vec{f}_b \ , 
\label{eq:mhd} 
\end{align} 
where $\vec{u}$ denotes the velocity
field, $\vec{b}$ the magnetic induction expressed in Alfv\'{e}n units, $\bm
\omega = \nabla \times \bm u$ the vorticity, $\bm j = \nabla \times \bm b$ the
current density, $P$ the pressure, $\vec{f}_u$ and $\vec{f}_b$ the external mechanical and electromagnetic forces, respectively.
Energy is dissipated at the small scales by the terms proportional to $\nu^+$
and $\eta^+$ and \blue{at} 
the large scales by $\nu^-$ and $\eta^-$. The indices $n,
m$ \blue{specify} 
the order of the Laplacian used. In order to obtain a large inertial
range, we chose $n = m = 4$. For all runs, we chose $\nu^+ = \eta^+$ and $\nu^-
= \eta^-$.  In the absence of forcing and dissipation, Eqs. \eqref{eq:mhd}
reduce to the ideal MHD equations, which have three conserved quantities: the
total energy $E = E_u + E_b = \frac{1}{2} \sum_{\bm k} (|\bm{u_k}|^2
+|\bm{b_k}|^2)$, the magnetic helicity $H_b = \sum_{\bm k}\bm{a_k} \cdot
\bm{b_{-k}}$, and the cross-helicity $H_c = \sum_{\bm k}\bm{u_k} \cdot
\bm{b_{-k}}$, where $\vec{a}$ denotes the vector potential of the magnetic
field.

The forces $\vf_u$ and $\vf_b$ are constructed from a randomized superposition
of eigenfunctions of the curl operator
\cite{Brandenburg01,Muelleretal12,MalapakaMuller13}, resulting \blue{in} Gaussian
distributed and $\delta(t)$-correlated forces whose helicities $\langle
\vf_{u,b} \cdot \nabla \times \vf_{u,b} \rangle$ and correlation $\langle \vf_u
\cdot \vf_b \rangle$ can be exactly controlled ($\avg{\cdot}$ indicates spatial
averages unless indicated otherwise). The specific random nature of the forces
ensures that at steady state the total energy input rate $\vep = \vep_u +
\vep_b = \langle \vu \cdot \vf_u\rangle + \langle \vb \cdot \vf_b\rangle
\propto |\vf_u|^2 + |\vf_b|^2$ is known \emph{a priori} \cite{Novikov65} with
$|\vf_u| = |\vf_b| = f_0$. In this case, $\vep$ can be used as a control
parameter.  Note that we choose to force both $\bm u$ and $\bm b$ with the same
forcing amplitude so that 
\blue{both quantities are dynamically important and $\bm b$ has a nonlinear feedback on the flow through the Lorenz force.} 

\blue{The use of an electromagnetic forcing is typical in studies that have
focused on the dynamics of magnetic helicity
\cite{Alexakisetal06,Muelleretal12,MalapakaMuller13} in contrast to dynamo
studies. Realistic analogues of such a forcing are, for example, the toroidal
current that is driven in tokamaks in order to generate a poloidal magnetic
field, and the force that is applied in electromagnetic pumps, which are driven
by means of a traveling magnetic field imposed by external coils. Note that
our intention in this study is not to simulate such a system but 
to further understand the dynamics of $H_b$, 
whose interpretation from dynamo simulations is ambiguous
because in these simulations $H_b$, and most importantly its mean flux, change sign in the inertial range 
of scales.
However, by forcing the
induction equation it is possible 
to maintain a 
sign-definite mean value of $H_b$,
which allows us to study the non-linear dynamics of magnetic
helicity across scales unambiguously.}

The forces are chosen such that the helicity of $\vf_u$ is negligible while $\vf_b$ is fully helical for all
simulations; and they are decorrelated, i.e., $\avg{\vf_u \cdot \vf_b} = 0$.
Thus, no $H_c$ and no kinetic helicity $H_u = \avg{\bm u \cdot \bm \omega}$ are
injected into the flow, while the injection of $H_b$ is maximal. 
\blue{Here, we exclude the injection of cross-helicity in the flow in order to avoid
introducing correlations between the velocity and the magnetic field which would
affect the cascade dynamics of the conserved quantities \cite{Linkmannetal16}.
Moreover, we avoid injecting kinetic helicity into the flow in order to exclude any
generation of mean magnetic helicity due to the presence of a strong component
of mean kinetic helicity in the flow.}
Finally, the initial \blue{Gaussian distributed} random 
magnetic and velocity fields are in equipartition
with energy spectra peaked at \blue{the forcing wave number} $k_f$ 
and zero helicities, i.e., $H_b = H_c = H_u
= 0$.  We should point out here that $H_c$ remains negligible in our flows,
however, $H_u$ increases with $H_u/(\avg{|\bm u|^2}\avg{|\bm \omega|^2})^{1/2}
\simeq 0.2$ at steady state. 

Equations \eqref{eq:mhd} are solved numerically using the standard
pseudospectral method, which ensures that $\nabla \cdot \vec{u} = 0$ and
$\nabla \cdot \vec{b} = 0$. Full dealiasing is achieved by the $2/3$-rule and
as a result the minimum and maximum wave numbers are $k_{box}$ = 1 and $k_{cut}$
= N/3, respectively, where $N$ is the number of grid points in each Cartesian
coordinate. Further details of the code can be found in
Refs.~\cite{Yoffe12,BereraLinkmann14}. 

Following previous studies \cite{BorueOrszag95,lamorgeseetal05}, a Reynolds number may be defined based on the control parameters
 of the problem as $Re_f \equiv u_f k_f^{1-2n}/\nu^+$ with $u_f \equiv \left(\vep/ k_f \right)^{1/3}$. 
In the following, $\vep$ and $\nu^+$ are adjusted such that $u_f$, and the ratio of $k_{cut}$ with the dissipation wave number 
$k_d \equiv (\vep/(\nu^+)^3)^{1/(6n-2)}$ remain the same for the different simulations. That is, $\vep/ k_f$ is kept constant between 
simulations while increasing the scale separation $k_fL$. This results in the same Reynolds number, 
which highlights that the only difference between the 
simulations is the scale separation between $k_f$ and $L$.
The numerical parameters of the simulations are given in Table~\ref{tbl:simulations}.

 \begin{table}[tbp]
 \begin{center}
 \begin{tabular}{*{7}{c}}
  $k_fL$ & $N$ & $Re_f$ & $\nu^+ = \eta^+$ & $\nu^- = \eta^-$ & $f_0$ & $T/t_f$ \\
  \hline
   10 & 128 & $7\times 10^3$ & $6.30 \times 10^{-12}$ & 0.05 & 1 & 320 \\ 
   20 & 256 & $7\times 10^3$ & $4.92 \times 10^{-14}$ & 0.05 & $\sqrt{2}$ & 130 \\ 
   40 & 512 & $7\times 10^3$ & $3.68 \times 10^{-16}$ & 0.05 & 2 & 100 \\ 
  \hline
  \end{tabular}
  \end{center}
\caption{Numerical parameters of the simulations. Note that $k_f$ denotes the forcing wavenumber, 
 $T$ the total runtime in simulation units, $f_0 = |\vf_u| = |\vf_b|$ the forcing amplitude 
 and $t_f \equiv (u_fk_{box})^{-1}$ a timescale defined based on the control parameters. 
All simulations are well resolved with $k_{cut}/k_d \geqslant 1.25$. 
 \label{tbl:simulations}
}
 \end{table}

The flux of total energy $\Pi_E(k)$ and magnetic helicity $\Pi_{H_b}(k)$ in Fourier space is given by 
\begin{align}
 \Pi_E(k) 
          & \equiv \avg{\vec{u_k}^<(\vec{\bm u} \cdot \nabla \vec{u}-\vec{\bm b} \cdot \nabla \vec{b})}
                  +\avg{\vec{b_k}^<(\vec{\bm u} \cdot \nabla \vec{b}-\vec{\bm b} \cdot \nabla \vec{u})}, \nonumber \\
 \Pi_{H_b}(k) & \equiv \avg{\vec{a_k}^<(\vec{\bm b} \cdot \nabla \vec{u}-\vec{\bm u} \cdot \nabla \vec{b})},
\end{align}
where the notation $g_k^<$ denotes the Fourier filtered field $g$ such that
only the modes satisfying $|\bm k| \leq k$ are kept \cite{Frisch95}. Negative
values of the fluxes imply inverse cascades while positive values imply forward
cascades.

The strength of forward and inverse cascades 
of a conserved quantity at steady state can be quantified by the rate of dissipation in the small and large scales, respectively, which we define as 
$\vep^\pm \equiv 2 \nu^\pm \left\langle\sum_{k\neq0} k^{\pm 2n} (|\vec{u_k}|^2 + |\vec{b_k}|^2) \right\rangle$,
where $\avg{\cdot}$ denotes a time-average in this case. Then, the total energy dissipation rate is given by $\vep = \vep^+ + \vep^-$.
A similar decomposition can be made for the dissipation rate of $H_b$ resulting in 
$\vep_{H_b} = \vep^-_{H_b} + \vep^+_{H_b}$ with 
$\vep_{H_b}^\pm \equiv 2 \nu^\pm \left\langle\sum_{k\neq0} k^{\pm 2n} (\vec{a_k}\cdot\vec{b_{-k}}) \right\rangle$.


Since we are interested in the behavior of \blue{the total energy and the magnetic helicity at large scales} 
we need to ensure that our simulations have been integrated for long enough times so that the
largest scales are in a statistically stationary state. This is satisfied when
$\vep^-$ reaches a saturated state. In what follows we analyze the data from
these saturated states.

\begin{figure}[!ht] 
 \centering
   \includegraphics[width=\columnwidth]{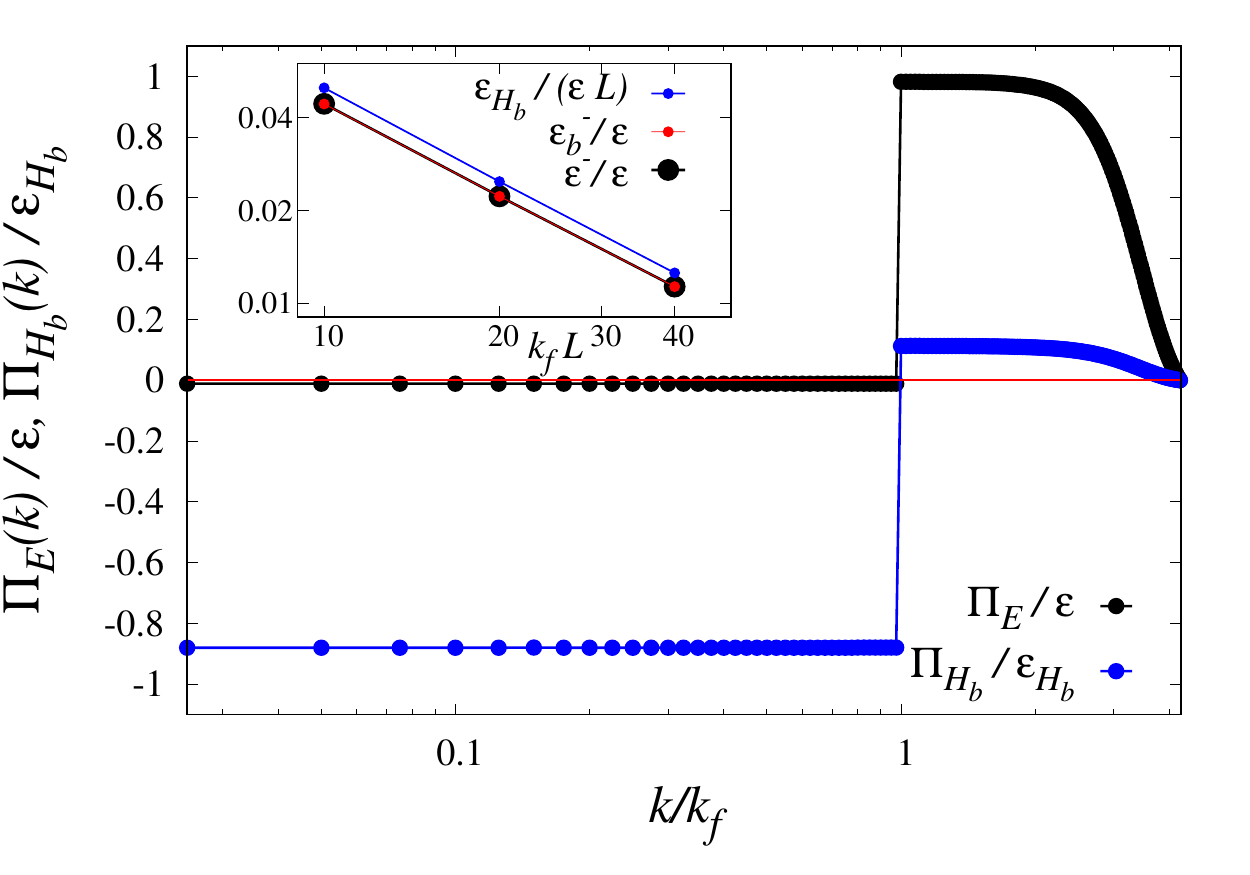}
\caption{ 
Fluxes of total energy and magnetic helicity normalized with the corresponding
dissipation rates for the run with $k_fL = 40$.
\blue{Black: normalized total energy flux $\Pi_E(k)/\vep$.
Blue (dark gray): normalized magnetic helicity flux $\Pi_{H_b}(k)/\vep_{H_b}$. }
The inset presents the scaling
$\vep^-/\vep \propto \vep_{H_b}/(\vep L) \propto (k_fL)^{-1}$.
\blue{
Black: $\vep^-/\vep$.
Red (light gray): $\vep_b^-/\vep$.
Blue (dark gray): $\vep_{H_b}/(\vep L)$.
}
}
    \label{fig:512_flux}
 \end{figure}

Absolute equilibrium (zero-flux) statistical mechanics have been used
effectively to suggest the directions of cascades (finite flux) of the ideal
conserved quantities across scales. \blue{In particular, it has been} 
shown
\cite{Frischetal75} that energy is transferred toward small scales (forward
cascade) while magnetic helicity is transferred toward large scales (inverse cascade).
\blue{In order to} check if these predictions are true in our flows 
we plot $\Pi_E(k)$ and
$\Pi_{H_b}(k)$ normalized by their dissipation rates (see Fig.~\ref{fig:512_flux}). 
The total energy has a forward cascade with $\Pi_E(k) > 0$ for $k \geq k_f$,
while $\Pi_E(k) \simeq 0$ for $k<k_f$.  The magnetic helicity, however, has a
dual cascade toward large and small scales, even though the injection of $H_b$
is maximal, with $\sim 90\%$ of $\Pi_{H_b}(k)$ being negative (i.e.,
$\vep^-_{H_b} \simeq 0.9 \vep_{H_b}$) at $k<k_f$.  From the inset of Fig.
\ref{fig:512_flux}, we observe that $\vep^- = \vep_b^- \propto \vep_{H_b}/L \propto (k_fL)^{-1}\vep$. 
This scaling implies that 
\blue{the ratio $\vep_{H_b}/(\vep L)$ determines}
the fraction of the total energy flux that proceeds
toward large scales.  This can be partly understood from the relation between
the injection rates of $H_b$ and $E_b$ due to the helical $\vf_b$ forcing, i.e.,
\beq
\vep_{H_b} = \langle \va \cdot \vf_b \rangle 
= k_f^{-1} \langle \va \cdot (\nabla \times \vf_b) \rangle
= k_f^{-1} \langle \vb \cdot \vf_b \rangle
= k_f^{-1} \vep_b.
\eeq 
Note that no matter how $\vep_{b}$ (and therefore $\vep$) may be varied with
$k_f$, \blue{one obtains} $\vep_b^-/\vep \propto (k_fL)^{-1}$, that is, 
for $k_fL \gg 1$ we expect $\Pi_{E_b}(k) = \vep_b^- \rightarrow 0$ at $k < k_f$  
and hence \blue{$\Pi_{E}(k) = \vep^- \rightarrow 0$ at $k < k_f$ since $\vep^- = \vep_b^-$ (see the inset of Fig. \ref{fig:512_flux})}. 
In other words, the inverse flux of total energy will become negligible once the
separation between the forcing scale and the largest scale of the system
becomes very large. This results in a weak energy input to sustain large-scale
magnetic fields.

\begin{figure}[!ht]
  \includegraphics[width=\columnwidth]{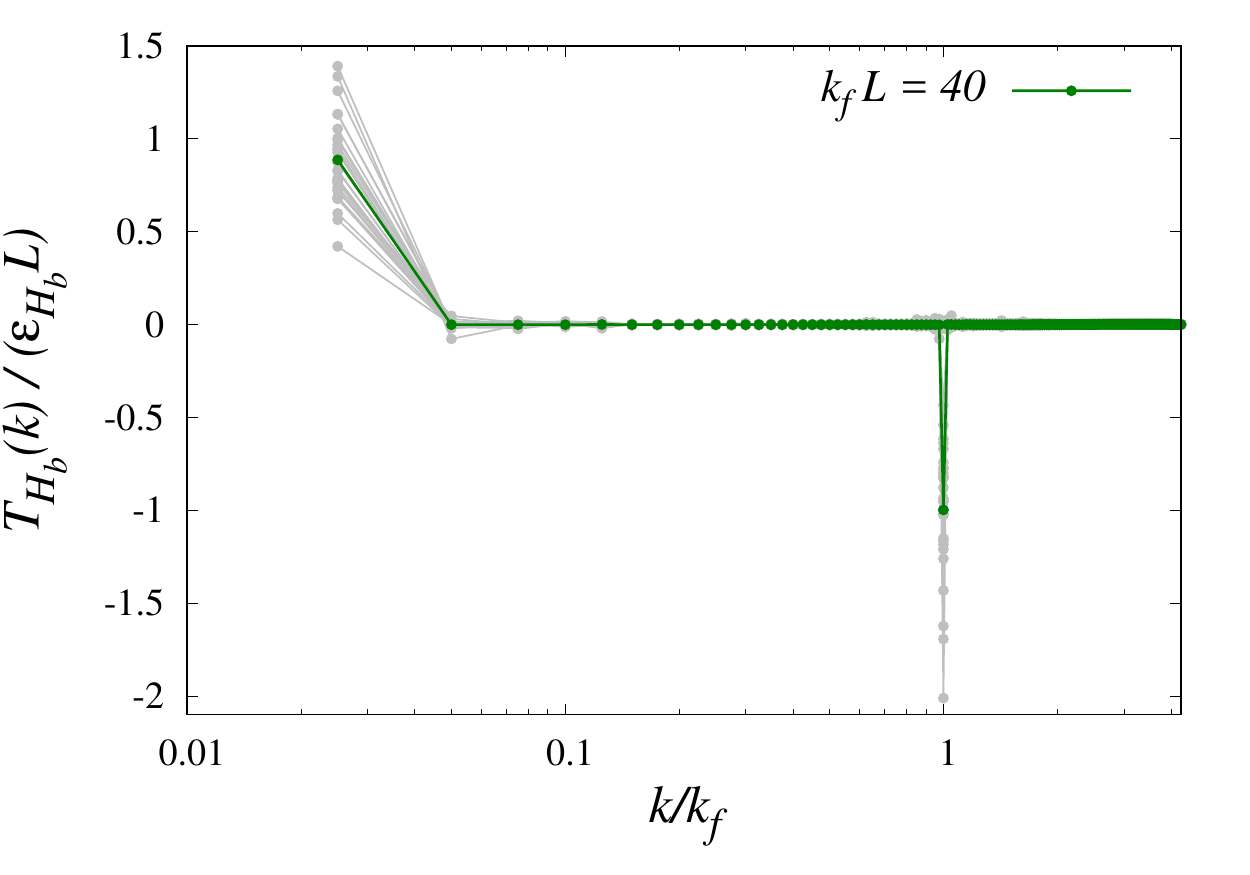}
  \caption{
Transfer spectra of magnetic helicity normalized with $\vep_{H_b}L$ for the flow with $k_fL = 40$. The time-averaged transfer $T_{H_b}(k)$ is indicated by the green (dark gray) curve while the gray curves indicate the instantaneous transfers $T_{H_b}(k,t)$.
}
 \label{fig:heltrans}
 \end{figure}

According to Fig. \ref{fig:512_flux}, $H_b$ has a constant negative flux at
$k<k_f$, which implies an inverse cascade. This is based on the idea that the
time-averaged transfer is zero in some intermediate wave-number range for flows
with large enough scale separation \cite{davidson04}.  We should point out here
that zero time-averaged transfer does not 
necessarily mean constant
{nonzero} $H_b$ transfer (i.e., no gain or loss of $H_b$ in a particular
wave number $k$), 
it can also mean zero $H_b$ transfer between modes. 
\blue{Considering} the instantaneous transfers,
\beq
T_{H_b}(k,t) \equiv \sum_{|\vk|=k} \sum_{\vp+\vq=\vk} \vb_{-\vk} \cdot (\vu_{\vp} \times \vb_{\vq}) \ ,
\eeq
at steady state (see Fig. \ref{fig:heltrans}), we see that $T_{H_b}(k,t) = 0$
for $5 < k < k_f$, implying not only vanishing average transfer but also no instantaneous transfer of $H_b$ into the large scales, 
apart from the fluctuations that are observed at the low $k$ modes. 
Therefore, this observation suggests a nonlocal transfer of magnetic helicity from the 
forcing scale to the largest scales of the magnetic field. 

\begin{figure}[!ht]
 \includegraphics[width=\columnwidth]{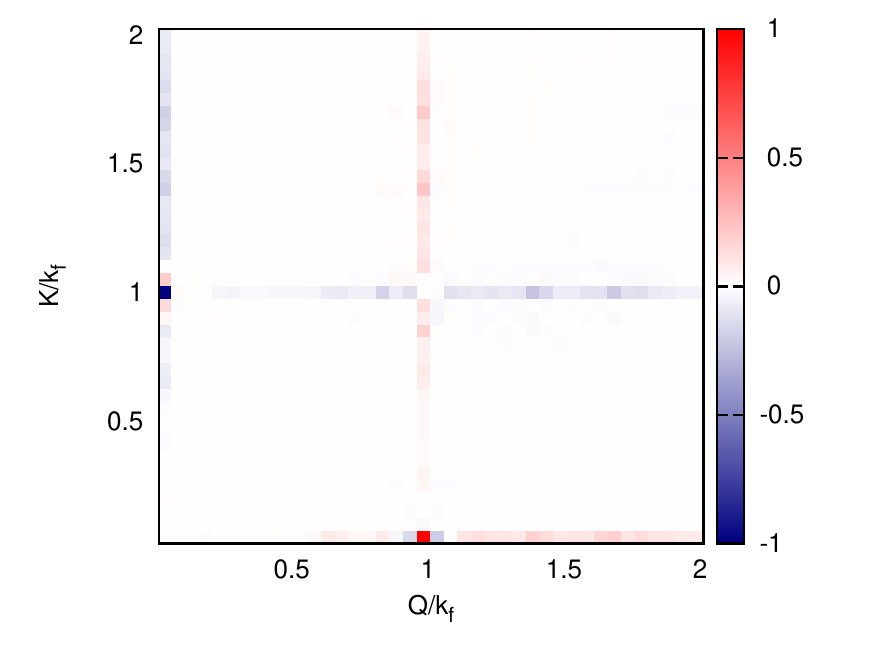}
 \caption{Instantaneous normalized magnetic helicity transfer $T_{H_b}(K,Q)/(\vep_{H_b}L)$ from shell $Q$ to shell $K$ rescaled with $k_f$.
}
\label{fig:shell-to-shell}
\end{figure}

To be precise on this statement, we analyze the shell-to-shell transfers
\cite{Alexakisetal06,Mininni11} of magnetic helicity, 
\beq 
T_{H_b}(K,Q) \equiv \int \vec{b}_K \cdot (\vec{u} \times \vec{b}_Q) d^3 \bm x 
\eeq 
from shell $Q$ to shell $K$. This transfer term conserves magnetic helicity, 
\blue{that is,} it does not
generate or destroy $H_b$, but it is responsible for the redistribution of $H_b$
across different scales. This is expressed by the fact that $T_{H_b}(K,Q)$ is
antisymmetric \blue{under the exchange of $K$ and $Q$}, i.e., $T_{H_b}(K,Q) = -T_{H_b}(Q,K)$, which is confirmed by Fig.~\ref{fig:shell-to-shell}.

Figure \ref{fig:shell-to-shell} shows that all transfers of $H_b$ between
wave numbers smaller than $k_f$ vanish apart from the transfer from the forced
shell $Q/k_f = 1$ to the smallest shell $K/k_f$.  Therefore, \blue{ 
the Fourier modes at
$k_f$ interact nonlocally with the Fourier modes at $k=1$} 
while the intermediate wave numbers $1 < k < k_f$ have
$T_{H_b}(K,Q) = 0$ and hence $\Pi_{H_b}(k) = 0$ in this range.  In other words,
the constant negative value of $\Pi_{H_b}(k)$ in the $k<k_f$ regime is a
manifestation of nonlocal transfers of $H_b$ from the forcing scale to the
largest scale of the magnetic field.

In summary, as we increase $k_fL$ we observe $\Pi_E(k) \rightarrow 0$ at
$k<k_f$ and $\Pi_{H_b}(k) = 0$ at $5 < k < k_f$.  Therefore, we expect the
large-scale flow to be described to a large extent from the predictions of the
zero-flux solutions given by absolute equilibrium statistical mechanics
\cite{Dallasetal15,Frisch95}.  To verify this we examine the scaling of our
spectra at scales larger than the forcing length scale but also sufficiently
smaller than the box size and we compare with \blue{the predictions from} 
absolute equilibrium theory, which we present below.

Following Ref.~\cite{Frischetal75}, we consider the Boltzmann-Gibbs distribution for
the truncated ideal 3D MHD equations (i.e., only Fourier modes $k_{min} \leq k
\leq k_{max}$ \blue{are} 
kept, $k_{max}$ being the truncation wave number) with
zero cross-helicity, $\mathcal P = Z^{-1} \exp(-\alpha E -\beta H_b)$, where $Z$
is the partition function of a Gaussian ensemble.  The coefficients $\alpha$
and $\beta$ are determined by the total energy and the magnetic helicity of the
system and can be interpreted as the inverse temperatures in the classical
thermodynamic equilibrium sense.  Using the discrete form of $E$ and $H_b$ we
can obtain the following expressions for their spectral densities at absolute
equilibrium:
\begin{align}
 E_u(k) &= \frac{4\pi}{\alpha}k^2 \ , \nonumber\\ 
 E_b(k) &= \frac{4\pi}{\alpha} \frac{k^2}{1 - (\frac{\beta}{\alpha k})^2}\ , \nonumber\\
 H_b(k) &= - \frac{4\pi\beta}{\alpha^2} \frac{1}{1 - (\frac{\beta}{\alpha k})^2}.
 \label{eq:statmech}
\end{align}
In order for $\mathcal P$ to be normalizable, i.e., the quadratic form $\alpha E
+\beta H_b$ to be positive definite, we need $\alpha > 0$ and $\alpha >
|\beta|/k_{min}$.  These spectra have a singularity at wavenumber $k_s \equiv
|\beta|/\alpha < k_{min}$ outside the validity of Eqs.~\eqref{eq:statmech}.

Equations \eqref{eq:statmech} suggest equipartition of kinetic energy across
scales while magnetic energy equipartition is only true for $k \gg k_s$. As $k
\rightarrow k_{min}$ the values of $E_b(k)$ and $H_b(k)$ diverge for values of
$k_s$ close to $k_{min}$. The region near $k_{min}$ has maximal helicity where
the total energy $E$ is dominated by $E_b$ and therefore $k_{min}|H_b| \simeq
E$.  This divergence of $H_b(k)$ at low $k$ (which is a conserved quantity
unlike $E_b$) is the indicator of the corresponding \blue{magnetic helicity transfer} 
toward large scales.

\begin{figure}[!ht]
  \includegraphics[width=\columnwidth]{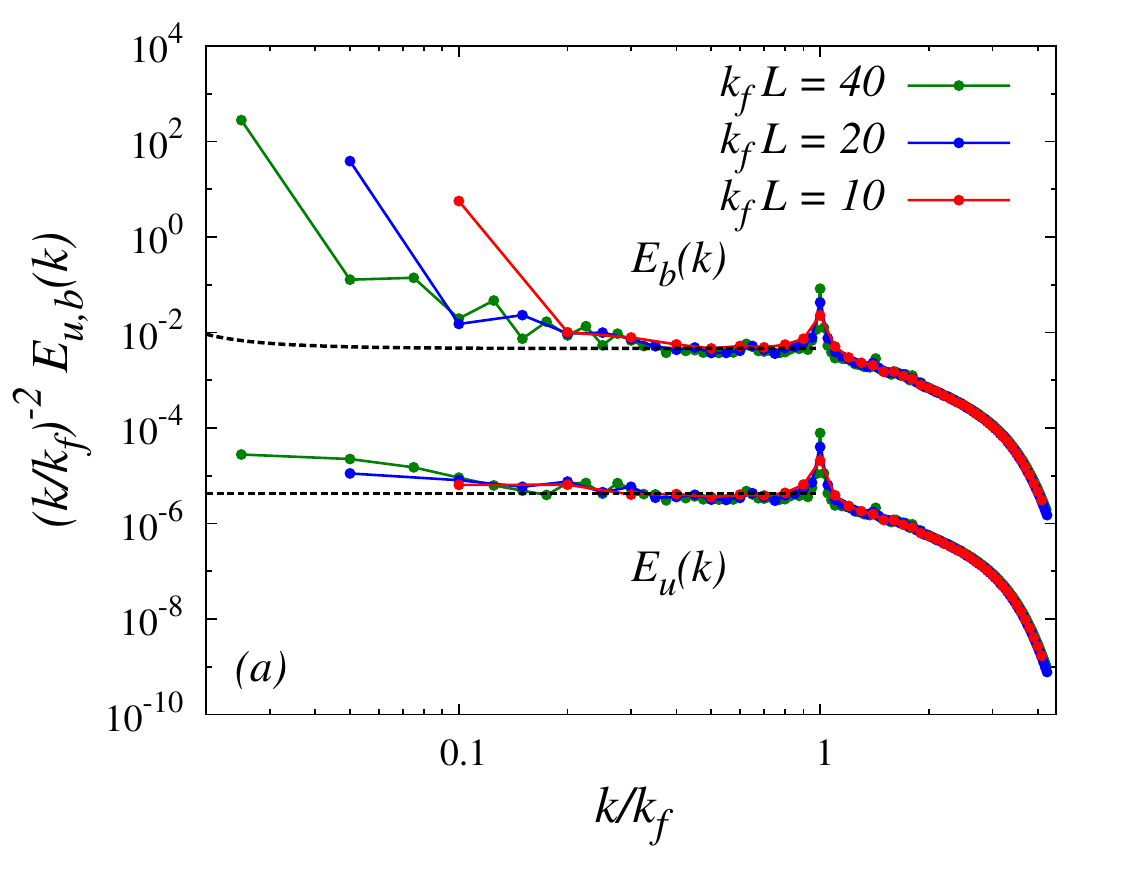}
  \includegraphics[width=\columnwidth]{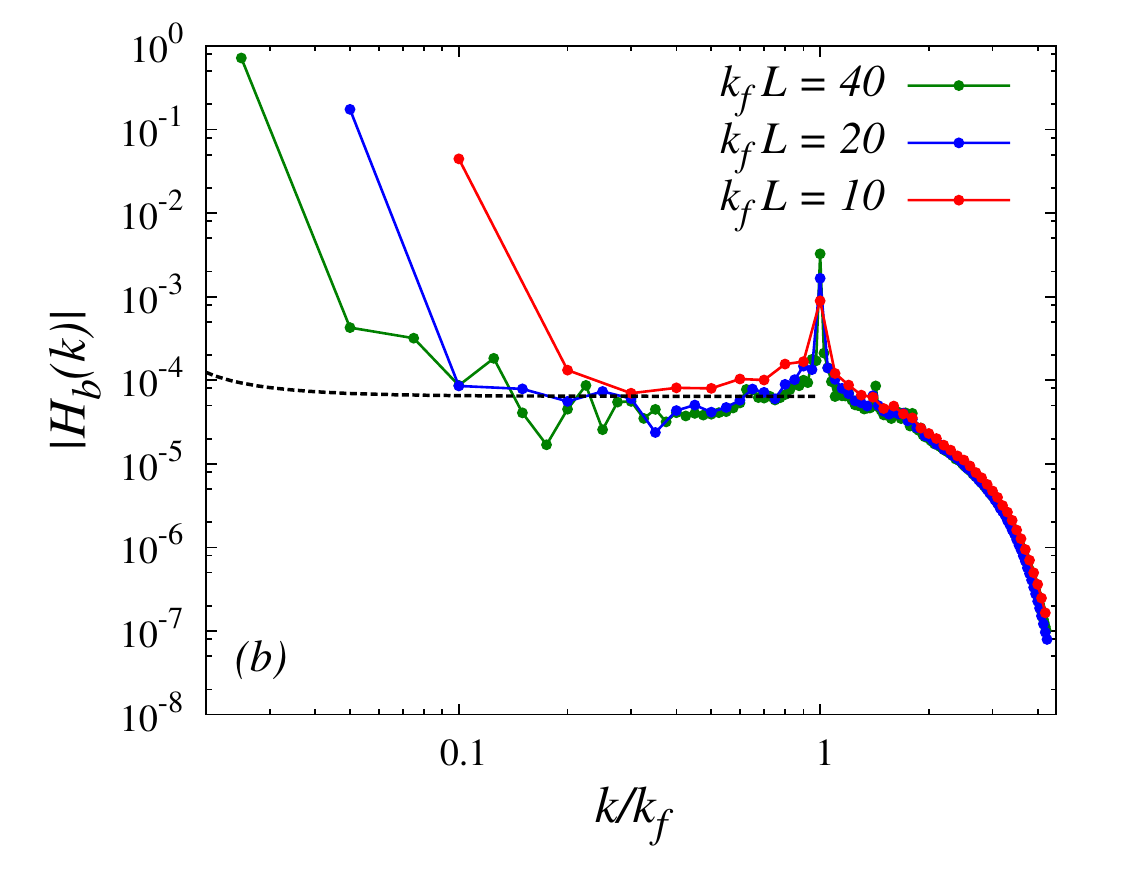}
 \caption{ 
(a) Magnetic and kinetic energy spectra 
compensated by $k^{-2}$, where $E_u(k)$ has been shifted by a factor of $10^{-3}$. 
(b) Absolute value of magnetic helicity spectra. The spectra are collapsed by rescaling with $k/k_f$. 
The dashed lines correspond to the predictions from absolute equilibria, i.e. Eqs.~\eqref{eq:statmech}. 
\blue{The green (light gray) curves refers to the simulation with $k_fL=40$, the blue (dark gray) to $k_fL=20$ 
and the red (gray) to $k_fL=10$.}
The inset shows the absolute value of the relative magnetic helicity 
$\rho_b(k)= H_b(k)/(\langle|\vec{a}_{\vk}|^2\rangle^{1/2}\langle|\vec{b}_{\vk}|^2\rangle^{1/2})$}.
\label{fig:specs}
\end{figure}

Figure \ref{fig:specs}(a) shows the magnetic and kinetic energy spectra
compensated with $k^{-2}$, with the $E_u$ spectra being shifted by a factor of
$10^{-3}$ for clarity. Note that $E_u(k)/E_b(k) \sim 1$ for wave numbers $k/k_f
\geq 0.3$.  The spectra here collapse by rescaling with $k/k_f$. 
Our data displays the scaling $E_u(k) \propto k^2$ at low wavenumbers $k<k_f$
with the range of validity increasing with $k_fL$.  The magnetic energy spectra
show a $E_b(k) \propto k^2$ scaling while the magnetic helicity [see
Fig.~\ref{fig:specs}(b)] show $H_b(k) \propto k^0$ for an intermediate range of
wavenumbers $k_{box} \ll k < k_f$.  These scalings are in agreement with the
predictions of the absolute equilibria for the truncated ideal MHD equations.
\blue{We point out}
that altering the hypodiffusive exponent $m$ does not 
affect the scaling of the spectra at \blue{the} large scales.
 
\blue{In order to compare the numerical results with the predictions of}
the absolute equilibrium theory, we have plotted 
Eqs.~\eqref{eq:statmech} as dashed lines in Fig.~\ref{fig:specs} using values of
$\alpha$ and $\beta$ obtained from a linear fit.  A measurable deviation from
the theory is observed as $k \rightarrow k_{box}$.  The amplitude of the
deviation is independent of the dissipation mechanism and only weakly dependent
on scale separation. The dashed lines in Fig.~\ref{fig:specs} predict that the
divergence of the spectra at $k_s = |\beta|/\alpha$ is expected at $k_s/k_f =
3.5 \cdot 10^{-4}$, which is beyond the expected validity of the absolute
equilibrium regime. Therefore, the deviation\blue{s} from the power-law scalings are
due to other possible reasons.  The most important reason affecting the
magnetic energy spectra is the minimal but still finite negative flux of $E_b$
at large scales. However, we expect better agreement of $E_b(k)$ with 
\blue{the corresponding absolute equilibrium prediction}
as scale separation increases because $\varepsilon_b^-/\varepsilon
\propto (k_fL)^{-1}$. 
The nonlocal transfers of magnetic helicity from the forcing scale to the
largest scales ($1 \leq k \leq 5$) of the flow is an obvious reason for
$H_b(k)$ to disagree with the equilibrium predictions.
Finally, for wave numbers close to $k_{box}$ the assumptions of isotropy used in
the derivation of Eqs. \eqref{eq:statmech} are not valid and deviations from
the isotropic result are expected for all the spectra in Fig. \ref{fig:specs}.

In the inset of Fig.~\ref{fig:specs}(b), we also plot the absolute value
of the relative magnetic helicity spectrum 
$\rho_b(k)= H_b(k)/(\langle|\vec{a}_{\vk}|^2\rangle^{1/2}\langle|\vec{b}_{\vk}|^2\rangle^{1/2})$
in order to quantify how far the flow is from a maximally helical state. It is
clear that the magnetic field becomes fully helical only at the largest scale of the flow 
while the rest of the scales have moderate values of $\rho_b(k)$. 

The magnetic field can also be sustained without forcing the induction equation, i.e., by dynamo action.
A recent numerical investigation \cite{Sadek16} of the kinematic dynamo with $k_fL \gg 1$
found that at scales larger than the forcing scale the kinetic energy scales as
$k^2$, in agreement with our results, while the magnetic energy \blue{scales} as $k^0$. We
expect that the scaling of the magnetic energy will change in the nonlinear
stage of the dynamo, where the Lorentz force is nonnegligible, and we
speculate that a $k^2$ spectrum will form in the saturated regime of the
dynamo. However, this is something that has to be confirmed.

In this paper, we investigated the dynamics of magnetic helicity focusing on
the large scales of MHD turbulence. We demonstrate that the inverse cascade of
magnetic helicity at steady state occurs nonlocally from the energy injection
scale into the largest scales of the flows we considered. By increasing scale
separation, we observe that the inverse energy transfer diminishes and hence no
magnetic helicity and almost no total energy is transferred to an intermediate
range of scales larger than the forcing length scale but also sufficiently
smaller than the box size. Therefore, we show that in this range of scales,
helical MHD turbulence is described to a large extent by the (zero flux)
absolute equilibrium spectra, with deviations expected at scales close to the
largest available scale of the system. 
\blue{Our results have direct implications for 
the understanding of the preservation of planetary and stellar magnetic fields on 
scales much larger than the size of these astrophysical objects.
}

Acknowledgements: We thank A. Berera for access to HPC resources and M. McKay
for technical help. This work has made use of the resources provided by ARCHER
\cite{archer}, made available through the Edinburgh Compute
and Data Facility (ECDF)\cite{ecdf}.  V.~D. acknowledges
support from the Royal Society and the British Academy of Sciences (Newton
International Fellowship, Grant No. NF140631).  M.~L. acknowledges support from the UK
Engineering and Physical Sciences Research Council (EP/K503034/1).  The
research leading to these results has received funding from the European
Union's Seventh Framework Programme (FP7/2007-2013) under Grant Agreement No.
339032.

\bibliography{refs}

\end{document}